\documentclass[newstyle,proceedings]{rmaa}
\usepackage{rmaacite}

\title{Building the bridge between Damped Ly$\alpha$ Absorbers
and Lyman-Break galaxies}
\author{
  J. P. U. Fynbo\altaffilmark{1},
  C. Ledoux\altaffilmark{2},
  P. M\o ller\altaffilmark{1},
  B. Thomsen\altaffilmark{3},
  I. Burud\altaffilmark{4},
  B. Leibundgut\altaffilmark{1}}

\altaffiltext{1}{ESO, Garching, Germany}
\altaffiltext{2}{ESO, Santiago, Chile}
\altaffiltext{3}{IFA, University of \AA rhus, Denmark}
\altaffiltext{4}{STScI, Baltimore, USA}

\fulladdresses{
\item J. P. U. Fynbo, P. M\o ller, B. Leibundgut: European Southern Observatory,
Karl-Schwarzschild-Stra\ss e 2, D-85748 Garching, Germany.
(jfynbo@eso.org,pmoller@eso.org,bleibund@eso.org)
\item C. Ledoux: European Southern Observatory,
Casilla 19001, Santiago 19, Chile.
(cledoux@eso.org)
\item B. Thomsen: Institute of Physics and Astronomy, University of \AA
rhus, Ny Munkegade, DK-8000 \AA rhus C, Denmark.
(bt@ifa.au.dk)
\item I. Burud: Space Telescope Science Institute, 3700 San Martin
Drive, Baltimore, MD-21218, USA.
(burud@stsci.edu)
}

\shortauthor{Fynbo et al.}
\shorttitle{The ``Building the Bridge'' survey}

\listofauthors{J. P. U. Fynbo, C. Ledoux, P. M\o ller, B. Thomsen, I.
Burud, B. Leibundgut}
\indexauthor{Fynbo, J. P. U.}

\ReceivedDate{}
\AcceptedDate{}
\SetYear{2002}

\abstract{
In 2000, we started the program ``Building the Bridge between
Damped Lyman-$\alpha$ Absorbers and Lyman-Break Galaxies: Ly$\alpha$
Selection of Galaxies'' at the European Southern Observatory's Very Large
Telescope (VLT). This project is an attempt to use the Ly$\alpha$ selection of
high-redshift galaxies to bridge the gap between absorption- and emission-line
selected galaxies by creating a large database of $z\approx 3$
galaxies belonging to the abundant population of faint (${\rm R}>25.5$)
galaxies associated to Damped Ly$\alpha$ Absorbers (DLAs). Here we present the
first part of our program, namely results from a deep Ly$\alpha$ study of
the field of the $z=2.85$ DLA toward the quasar Q\,2138$-$4427.
}

\keywords{Galaxies: high-redshift}
\begin{document}
\maketitle

\section{Introduction}

During the last decade, the observational study of high-redshift galaxies went
through a revolution due to the advent of 8--10\,m class telescopes and
their instrumentations, allowing spectroscopic redshifts to be measured
for galaxies as faint as ${\rm R}\sim 25$. Most of this progress, in terms of
sample size, came from the study of Lyman-Break Galaxies (LBGs) selected
by the sharp drop in the continuum of star-forming galaxies at the Lyman
limit (Steidel et al. 1996; Fontana et al. 2000). In addition, the study of
the dust and metal contents of high-redshift Damped Ly$\alpha$ Absorbers
(DLAs) provides an independent look at the properties
of high-redshift galaxies (see Pettini et al. 1997; Ledoux et al. 2002 and
refs. therein). However, the number of galaxy counterparts of DLAs detected in
emission has remained low, with only a handful of successes
(see M\o ller et al. 2002 and refs. therein), implying that the bulk of the
DLA galaxy population is significantly fainter than LBGs. This is because LBG
samples are continuum-flux limited and the current spectroscopic limit
of ${\rm R}=25.5$ in (ground-based) samples is not deep enough to reach the
level of galaxies probed by DLAs (Fynbo et al. 1999).

\section{The ``Building the Bridge'' survey}
In 2000, we started the program ``Building the Bridge between
Damped Lyman-$\alpha$ Absorbers and Lyman-Break Galaxies: Ly$\alpha$
Selection of Galaxies'' at the
VLT ($+$FORS). This project is an attempt to use the Ly$\alpha$ selection to
bridge the gap between absorption- and emission-line selected galaxies
by creating a large database of $z\approx 3$ galaxies belonging to the
abundant population of faint (${\rm R}>25.5$) galaxies associated to DLAs
(Fynbo et al. 1999; Haehnelt et al. 2000). While targetting the fields of QSO
metal-rich absorption line systems at $z\approx 3$, our main goal is not to
detect the galaxy counterparts of the absorbers but to anchor our fields
to previously known structures at the target redshift, therefore minimizing
the risk of observing a void. We choose to use Ly$\alpha$ selection as it has
been shown to be an efficient way to probe the faint end of the
galaxy luminosity function of high-$z$ galaxies (see M\o ller et al. 1993;
Francis et al. 1995; Cowie \& Hu 1998; Kudritzki et al. 2000;
Pentericci et al. 2000, Fynbo et al. 2001 for examples
of spectroscopically confirmed samples).

\begin{figure}[!t]
\includegraphics[width=\columnwidth]{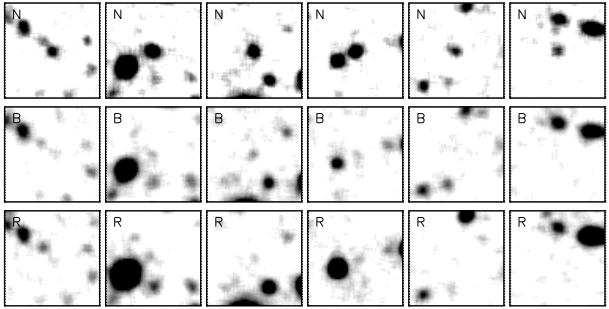}
\caption{
Six examples of candidate $z=2.85$ Ly$\alpha$ emitters in the field
of Q\,2128$-$4427. The size of each image is $12\times 12$ arcsec$^2$.
The upper rows are the narrow-band images of the candidates, the middle rows
the B-band images and the bottom ones the R-band images.
Several ($\sim 20$\%) of our candidates remain undetected in the broad bands
despite $5\sigma$ detection limits of ${\rm B}({\rm AB})=27.0$
and ${\rm R}({\rm AB})=26.4$.
}
\label{fig1}
\end{figure}

\section{First results}
So far, the observations of the field of the $z=2.85$ DLA towards the
quasar Q\,2138$-$4427 are complete. In this field, we detect 35 candidate
Ly$\alpha$ emitters probing a redshift range $\Delta z=0.053$. Six examples
are shown in Fig.~\ref{fig1}. The inferred density of candidates is about
15 arcmin$^{-2}$ per unit redshift to a $5\sigma$ flux limit
of $6\times 10^{-18}$ erg s$^{-1}$ cm$^{-2}$, a surface density which is about
15 times larger than that for LBGs (Steidel et al. 1996). This is consistent
with the fact that $\sim 90$\% of our candidates are fainter than
the ${\rm R}=25.5$ spectroscopic limit for LBGs. Hence, we have
already bridged at least some of the gap between LBGs and DLAs.

\begin{figure}[!t]
\includegraphics[width=\columnwidth]{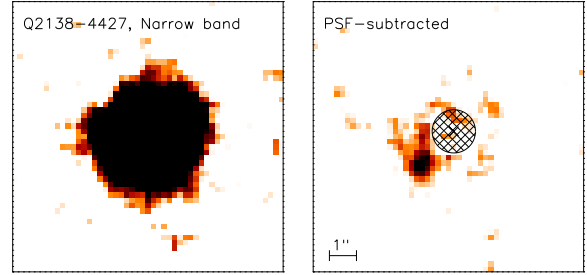}
\caption{
{\it Left panel:} A $10\times 10$ arcsec$^2$ narrow-band image centred on the
quasar Q\,2128$-$4427. {\it Right panel:} The same region after PSF
subtraction of the QSO light showing excess emission 1.4 arcsec SE of the
QSO position. The hashed area indicates the region where the residuals from
the PSF subtraction are large. With upcoming spectroscopic observations with
FORS at the VLT, we will be able to determine if this candidate is actually
the DLA galaxy counterpart.
}
\label{fig2}
\end{figure}

The glare of the quasar is reduced by 1.5 magnitudes in the narrow-band
image due to the strong Damped Ly$\alpha$ Absorption line. This makes it much
easier to detect the faint emission from a possible galaxy counterpart of
the DLA close to the line of sight. After PSF subtraction, we detect an
extended source at an impact parameter of only 1.4 arcsec from the QSO in
the narrow-band image. This is a prime candidate for the DLA
galaxy (see Fig.~\ref{fig2}).

Spectroscopic follow-up of this first sample of candidate Ly$\alpha$ emitters
is under way.

\adjustfinalcols

\end{document}